\documentclass[12pt, preprint]{aastex}

\usepackage{graphicx}
\usepackage{url}
\usepackage{natbib}
\usepackage{epsfig}
\usepackage{amsmath}
\usepackage{amssymb}
\usepackage{color}
\usepackage{subfig}

\newcommand{\rmd}{\textrm{d}}

\newcommand{\Jbar}{\bar{J}}

\usepackage{ulem} 

\newcommand{\Fermi}{\emph{Fermi-LAT}}

\newcommand{\Unit}[1]{\,\mathrm{#1}}

\journal{Astroparticle Physics}

\begin{document}

%------------------------------------------------------------
%                           TITLE
%------------------------------------------------------------
\title{Prospects for a Dark Matter annihilation signal towards the Sagittarius dwarf galaxy with ground based Cherenkov telescopes}
%------------------------------------------------------------
%                           AUTHORS & labs
%------------------------------------------------------------

 \author{A.~Viana\altaffilmark{1}, M.C.~Medina\altaffilmark{1}, J. Pe{\~n}arrubia\altaffilmark{2}, P.~Brun\altaffilmark{1}, J.F.~Glicenstein\altaffilmark{1}, K.~Kosack\altaffilmark{1}, E.~Moulin\altaffilmark{1},
 M.~Naumann-Godo\altaffilmark{1}, B.~Peyaud\altaffilmark{1}}
\altaffiltext{1}{IRFU/DSM, CEA Saclay, F-91191 Gif-sur-Yvette, Cedex, France}
\altaffiltext{2}{Institute of Astronomy, University of Cambridge, Cambridge, CB3 0HA, United Kingdom}

\begin{abstract}
  Dwarf galaxies are widely believed to be among the best targets for
  indirect dark matter searches using high-energy gamma rays; and
  indeed gamma-ray emission from these objects has long been a subject
  of detailed study for ground-based atmospheric Cherenkov telescopes.
  Here, we update current exclusion limits obtained on the closest
  dwarf, the Sagittarius dwarf galaxy, in light of recent realistic
  dark matter halo models. The constraints on the velocity-weighted
  annihilation cross section of the dark matter particle are of a few
  10$^{-23}$ cm$^{3}$s$^{-1}$ in the TeV energy range for a 50~h
  exposure. The limits are extrapolated to the sensitivities of future
  Cherenkov Telescope Arrays.  For 200~h of observation time, the
  sensitivity at 95\% C.L. reaches 10$^{-25}$ cm$^{3}$s$^{-1}$. Possible astrophysical backgrounds from gamma-ray sources dissembled in Sagittarius dwarf are studied. It is shown that
  with long-enough observation times, gamma-ray background from
  millisecond pulsars in a globular cluster contained within
  Sagittarius dwarf may limit the sensitivity to dark matter
  annihilations.
\end{abstract}

%------------------------------------------------------------
%                          KEYWORDS
%------------------------------------------------------------


%------------------------------------------------------------
%                          TEXT
%-----------------------------------------------------------

\newpage
\newpage
\section{Introduction \label{sec:part1}}

Dark matter (DM) plays a key role in the dynamics of a large class of
astrophysical systems in the Universe.  Though halos of dark matter
are predicted to exist around all galaxies, dwarf spheroidal galaxies
in particular are ideal targets for DM annihilation searches because:
(i) their stellar dynamics show that they are among the most
DM-dominated objects in the Universe; (ii) due to the lack of recent
star formation activity, their environment is relatively quiet in
terms of background astrophysical gamma-ray emission ; (iii) many of
them lie at distances below 100 kpc from the Galactic Center.

The search for secondary gamma-rays from annihilations of dark matter
particles is a powerful indirect detection technique because
gamma-rays do not suffer from propagation effects, the gamma-ray
signal should be proportional to the square of the DM density, and
characteristic features such as lines or steps may be present in the
energy spectrum at these energies~\citep{2000RPPh...63..793B,Bringmann:2007nk}.  Imaging atmospheric
Cherenkov telescopes (IACTs) such as HESS~(\citeyear{hess}),
MAGIC~(\citeyear{magic}) and VERITAS~(\citeyear{veritas}), are
particularly well suited to deep searches of targeted objects because
of their large effective areas ($\sim10^5\Unit{m^2}$ above 100 GeV).
However, since IACTs are multipurpose astrophysical experiments and
have a short duty cycle ($\sim$1000 hours/year), the observation time dedicated to these
objects is typically limited to tens of hours per year.

Since the flux of the expected gamma-ray signal is inversely
proportional to the square of distance, one would expect the best
dwarf spheroidal target to be the nearest one.  However such dwarfs
are also the closest to the Galactic Center and experience the tidal
effect of the Milky Way.  Recently, it has been shown that one could
take advantage of this effect to trace back the evolution history of
the object~\citep{2008ApJ...673..226P}.  During the orbital
motion of a dwarf galaxy, multiple crossings of the dwarf galaxy
through the galactic disc of the Milky Way give rise to the formation
of tidal streams, a careful study of which allows to one infer the
gravitational potential of the dwarf galaxy.

In the case of the Sagittarius Dwarf galaxy (SgrDw), the tidal streams
have been detected with multiple tracer
populations~\citep{Yanny:2000ty,Vivas:2001dn,Watkins:2009im,1998ARA&A..36..435M,Majewski:1999sj,MartinezDelgado:2000id,Newberg:2001sx,Majewski:2003ux}
and have been used to derive the DM halo potential.  Furthermore,
measurements of stars within SgrDw and the luminosity of its core and
surrounding debris, allows the estimate of the DM content
\emph{prior} to tidal
disruption~\citep{NiedersteOstholt:2010hg,2010MNRAS.408L..26P}.  Other
peculiar features of SgrDw include the presence of the M54 globular
cluster coincident in position with its center of
gravity~\citep{Ibata:1994fv}, and hints for the presence of a central
Intermediate Mass Black Hole~\citep{2009ApJ...699L.169I} (IMBH). The latter
point is supported by the observation of a deviation from a flat
behavior in the surface brightness density profile towards the center
of the object.

Constraints on a DM annihilation signal towards SgrDw, Canis Major,
Sculptor and Carina have been reported by
HESS~\citep{2008APh....29...55A,2009ApJ...691..175A,aharonian:2010zzt},
towards Draco, Willman 1 and Segue 1 by MAGIC~\citep{Albert:2007xg,Aliu:2008ny,Aleksic:2011jx}
and towards Draco, Ursa Minor, Bo\"{o}tes 1 and Willman 1 by
VERITAS~\citep{Acciari:2010pja}, and towards Draco and Ursa Minor by Whipple~\citep{Wood:2008hx}.
Because of its location in the Southern hemisphere, HESS is better
suited for observations of SgrDw with respect to other
currently-operating IACTs. Observations with the Fermi Large Area Telescope (\Fermi, a
space-based telescope sensitive to gamma-rays between 20 MeV and 300 GeV), are also
well-suited due to its large duty cycle and wide field-of-view, though
the energy range probed is lower than that of IACTs.  The
\Fermi\ collaboration put strong constraints in the GeV DM
mass range on dwarf spheroidal galaxy satellites of the Milky
Way~\citep{2010ApJ...712..147A,collaboration:2011wa}. However, their study is restricted to
high galactic latitude ($\rm |b|>30^{\circ}$) objects to avoid
systematic contamination from galactic diffuse gamma-ray emission, and
therefore no constraints on the measured flux in the direction of
SgrDw have yet been published using \Fermi\ data.

In this paper, the current constraints on a DM annihilation signal
towards SgrDw are reassessed in light of more realistic DM halo models
than previously used~\citep{Evans:2003sc,2008APh....29...55A}.  The
sensitivity of the future generation of IACTs, \textit{i.e.}
CTA (Cherenkov Telescope Array, ~\citeyear{Consortium:2010bc}), is used to evaluate its potential for
the detection of a DM annihilation signal. The CTA design-study
sensitivity is used to investigate possible conventional gamma-ray
emission, \textit{e.g.} to the population of millisecond pulsars (MSP)
in the globular cluster M54 at the center of SgrDw, or from the jet of
a hypothetical central IMBH. It is shown that such standard
astrophysical signals may limit the sensitivity to DM annihilations
with CTA in case of long observation times, eventually requiring the
modelling and subtraction of these astrophysical components.

The paper is structured as follows: Section~\ref{sec:searches} is
dedicated to the description of current and future instruments as well
as the calculation of the sensitivity to DM signals.  In
Section~\ref{sec:modelling}, the modelling of the DM halo of SgrDw is
described together with the astrophysical contribution to the DM
flux. In the absence of an astrophysical gamma-ray background,
exclusion limits on the velocity-weighted annihilation cross section of
DM are derived in
Section~\ref{sec:limits}. Section~\ref{sec:backgrounds} deals with the
estimate of the gamma-ray emission from the MSP population and the
IMBH candidate of M54. Section 6 is devoted to the summary.

\section{Dark matter searches with IACTs \label{sec:searches}}

\subsection{Current and future instruments}\label{sec:currentinst}

The present generation of IACTs (HESS, MAGIC and VERITAS) consists of
multiple-telescope arrays detecting very high energy (VHE, E$_{\gamma}$ $\gtrsim$ 100 GeV) gamma-rays.
The stereoscopic view of extensive air showers generated in the
atmosphere by VHE gamma-rays allows these instruments to accurately
reconstruct the direction and the energy of the primary gamma-ray. The
angular resolution reaches 0.1$^{\circ}$ per gamma-ray event and the
point source sensitivity is about a few percent of the Crab Nebula
flux above $100\Unit{GeV}$~\citep[see, for
instance,][]{2006A&A...457..899A}.

The plan for the next generation of IACTs, the Cherenkov Telescope
Array (CTA, ~\citeyear{Consortium:2010bc}), involves building two large
arrays, one in each hemisphere, with  an order of magnitude more
telescopes than current instruments.  This future instrument is
expected to increase the flux sensitivity by a factor of 10 compared
to current instruments, and enlarge the accessible energy range both
towards the lower and higher energies. Based on the current CTA design
study, a factor of about ten in effective area and a factor of two better in hadron rejection are expected. In this study,
the estimated CTA effective area at the trigger level (before offline
gamma-hadron separation) is extracted from \cite{TesePaz}. In order to
mimic the effect of the analysis event selection, the effective area
values for energies from $\sim$100 GeV down to $\sim$20 GeV are
realistically lowered. The effective area then decreases from
$\sim10^6\Unit{m^2}$ at 200 GeV down to $\sim 10^3\Unit{m^2}$ at about
20 GeV.

\subsection{Sensitivity calculation and background estimates}\label{sec:sensitivity}

The sensitivity for IACTs is calculated by comparing the number of
events expected from an assumed gamma-ray emission scenario with the
expected level of background events. In the case of DM searches, the
assumed emission is from the annihilation of DM particles of mass $m$ in the halo
of the host galaxy, the differential gamma-ray flux of which is given by:

\begin{equation}
\label{eqnp}
\frac{\rmd\Phi(\Delta\Omega,E_{\gamma})}{\rmd E_{\gamma}}\,=\frac{1}{8\pi}\,\underbrace{\frac{\langle
    \sigma v\rangle}{m^2}\,\frac{\rmd N_{\gamma}}{\rmd E_{\gamma}}}_{\rm Particle\,
  Physics}\,\times\,\underbrace{\bar{J}(\Delta\Omega)\Delta\Omega}_{\rm Astrophysics} \, ,
\end{equation}
where $\langle\sigma v\rangle$ is the velocity-weighted annihilation cross-section and $\rmd N_{\gamma}/\rmd E_{\gamma}$ the photon spectrum per annihilation. The astrophysical factor is defined as
\begin{equation}
\Jbar(\Delta\Omega) = \frac{1}{\Delta \Omega} \int_{\Delta \Omega} \rmd\Omega \int_{\rm LOS} \rho^2[r(s)] \rmd s \, .
\label{jbar}
\end{equation}
When treating the self-annihilation of DM particles, this factor scales
with the squared density of DM, $\rho^2$, over the whole observation
cone. The integral is then taken along the line of sight (LOS) and
inside the solid angle $\Delta\Omega$. The solid angle is chosen as the
angular resolution for point-like
searches. The number of expected signal events can be calculated by:
\begin{equation}
N_{\gamma} = T_{\rm obs} \int_{0}^{\infty} A_{\rm eff}(E_{\gamma})\, \frac{\rmd \Phi_{\gamma}}{\rmd E_{\gamma}} \, \rmd E_{\gamma} \, ,
\label{ngamma}
\end{equation}
where $T_\mathrm{obs}$ is the observation time, and
$A_\mathrm{eff}(E_{\gamma})$ is the effective area of the detector as
a function of the gamma-ray energy. In the case where the background is not measured experimentally, it can still be estimated assuming that the background consists of misidentified hadron showers. The estimate of the expected number
of background events in the signal region can be determined using the
following expression~\citep[see][]{1998APh.....9..137B}:
\begin{equation}
\frac{\rmd^2 \Phi_{\rm had}}{\rmd\Omega \rmd E_{\gamma}} = 8.2 \times 10^{-8} \epsilon_{\rm had} \left(\frac{E_{\gamma}}{\rm 1 TeV} \right)^{-2.7} [{\rm TeV^{-1} cm^{-2} s^{-1} sr^{-1}}] \, ,
\label{back}
\end{equation}
where $\epsilon_{\rm had}$ is the hadron detection efficiency. To take
into account the performance of the future IACTs the hadron rejection
is taken at the level of 90$\%$, which corresponds to $\epsilon_{\rm
  had} = 0.1$. This parametrisation gives remarkable agreement with CTA background simulations~\citep{dipierro}.

In case of no gamma-ray signal, a limit on the number of gamma rays at 95\% confidence level (C.L.), $N^{\rm 95\% C.L.}_{\gamma}$, can be calculated using the method of \citet{Rolke:2004mj}. In what follows two cases are considered. In the case of current IACTs, the $N^{\rm 95\% C.L.}_{\gamma}$ calculation uses the numbers of gamma-ray and background events extracted from 11h H.E.S.S. measurements ~\citep{2008APh....29...55A}. The projected $N^{\rm 95\% C.L.}_{\gamma}$ for 50 h observation time is obtained by extrapolating both the numbers of gamma-ray and background events from 11 h to 50 h. In the case of 95\% C.L. sensitivity calculations, $N^{\rm 95\% C.L.}_{\gamma}$ is calculated assuming the background-only hypothesis. For the H.E.S.S. sensitivity the number of background events is taken from the extrapolation at 50 h of observation. For the CTA sensitivity, the number of background events is calculated by integrating the background event flux given in Eq.~(\ref{back}) after multiplication by the effective area of the detector and the observation time. $N^{\rm 95\% C.L.}_{\gamma}$ is then calculated using five off regions.

Replacing Eq.~(\ref{eqnp}) in Eq.~(\ref{ngamma}), the DM sensitivity can be then expressed in terms of the remaining particle physics parameters, $\langle \sigma v \rangle$, $m$ and $dN/dE_{\gamma}$. The 95\% C.L. limit on the velocity-weighted annihilation cross section is given by the following expression:

\begin{equation}
\left\langle \sigma v \right\rangle_{\rm min}^{\rm 95\% C.L.} = \frac{8\pi}{\Jbar(\Delta \Omega)\Delta \Omega} \times \frac{m^{2}\,  N_{\gamma}^{\rm 95\% C.L.}}{T_{\rm obs}\, \int_{0}^{m} A_{\rm eff}(E_{\gamma}) \, \frac{\rmd N_{\gamma}}{\rmd E_{\gamma}}(E_\gamma) \, \rmd E_{\gamma}} \, .
\end{equation}

\section{Modelling the Sagittarius dwarf dark matter halo \label{sec:modelling}}
The Sagittarius dwarf (Sgrdw) is the only satellite galaxy in the MW that shows clear evidence of ongoing tidal mass stripping~\citep{Ibata:2000ys} in the form of an associated tidal stream~\citep{1998ApJ...508L..55M,Majewski:1999sj,Majewski:2003ux,MartinezDelgado:2000id,MartinezDelgado:2003qy,Belokurov:2006ms,Watkins:2009im}. This galaxy is currently located at a close distance from the MW centre~\citep[$\approx 17$ kpc;][]{1998ApJ...508L..55M}. Indeed, it underwent its last perigalacticon passage only 17 Myr ago~\citep{Law:2010pe,Penarrubia:2008mu}, which is a relatively short time compared with its internal dynamical time $t_{\rm dyn}=R_c/\sigma_0\approx 47$ Myr, where $R_c$ is the galaxy core radius and $\sigma_0$ the central velocity dispersion~\citep{1998ARA&A..36..435M}.

The proximity of the Sgrdw to the Milky Way plus the fact that this galaxy is shedding stars to tides complicates its dynamical modelling in a number of ways. On the one hand, the distribution of dark matter and stars has been clearly altered from its original configuration by tidal mass stripping. Given that the actual amount of stars and dark matter in the tidal tails is unknown~\citep{NiedersteOstholt:2010hg}, the original mass, luminosity and size of the Sgrdw remain fairly uncertain quantities. On the other hand, the assumption of dynamical equilibrium may not be adequate, specially in the outskirts of the galaxy where the population of unbound stars may dominate in number over that of bound members~\citep{Penarrubia:2008mu}.

These difficulties have not deterred a large body of theoretical work devoted to uncover the actual content and distribution of DM in the Sgrdw. To date these efforts have focused on (i) analytical models of the dynamical properties of the remnant core and (ii) N-body simulations that aim to reproduce the spatial and kinematical distribution of the tidal tails.

The simplest analytical models assume dynamical equilibrium and adopt a cosmologically-motivated halo density profile to describe the kinematics of individual stars
\begin{equation}
\rho_{\rm NFW}(r) =
\frac{\rho_{\rm s}}{(r/r_{\rm s})(1+ r/r_{\rm s})^2} \, ,
\label{eq:nfw}
\end{equation}
where $r_{\rm s}$ is a scale radius and $\rho_{\rm s}$ is a
characteristic density (Navarro, Frenk \& White 1996, hereafter NFW). Note that this profile diverges at small radii as $\rho\propto r^{-1}$, which is typically referred as a dark matter ``cusp''. It was shown in \cite{2008ApJ...673..226P} that the tightly bound dark matter cusp is more resilient to disruption than the more loosely bound stellar cored profile, which can be accurately described with a King (\citeyear{King:1966fn}) profile~\citep{1998ARA&A..36..435M}, and that tidal stripping does not change the inner profile of DM haloes.

Assuming that the external tidal field does not influence the kinematics of stars that locate the central regions of the dwarf, and ignoring the effects of tidal stripping on the outer ($r\gg r_s$) dark matter halo profile, one can use the Jeans equations to search the DM halo parameters that best fit the stellar central velocity dispersion for a observed King "core" radius of this object. The King-NFW degeneracy gives rise to a family of NFW halo models which can reproduce the stellar dynamics~\citep{2008ApJ...672..904P}. One way to break this degeneracy is using the relationship between the virial mass and concentration found in cosmological N-body simulations~\citep[see for instance, ][]{2001MNRAS.321..559B}. Using this procedure on the SDSS survey data provides a value of $r_s =$ 1.3 kpc. Considering the scatter on the relationship between virial mass and concentration, the 2$\sigma$ error on $r_s$ is found to be $\sim$0.2 kpc. This correspond to the family of models with $\rho_s$ spanning from $7.5\rm \times 10^{-3}$ to $1.3\rm \times 10^{-2}$ $M_{\odot} pc^{-3}$. In Table~\ref{tab:jbarvalue} we show the results of our fits together with the astrophysical factors $\Jbar$ for different solid angles $\Delta\Omega$. Taking into account the error on the halo profile parameters the value of the astrophysical factor can vary by a factor of 2. Interestingly, an independent analysis by \cite{2010ApJ...725.1516L} provides similar values for these parameters. In this case the astrophysical factors are found to be of a few higher than the ones presented here.

However, numerical N-body models that aim to describe the observed structural and kinematical distributions of stars in the tidal tails as well as the remnant core provide a more consistent approach to the dynamical analysis of the Sgrdw. Yet, most of the existing N-body models of this galaxy assume for simplicity that dark matter and stars share the same spatial distribution (the so-called ``mass-follows-light models''), an assumption that is not supported by detailed kinematic data of dwarf spheroidal galaxies (e.g.~\cite{Walker:2009zp}).
The only exception to date corresponds to recent N-body models constructed by \cite{2010MNRAS.408L..26P}, who explore the possibility that the Sgrdw may have originally been a rotating galaxy.
In these models the galaxy is composed of an exponential stellar disk embedded in an extended DM halo. The DM density profile is taken as a cored isothermal (ISO)
profile
\begin{equation}
\rho_{\rm ISO}(r)= \frac{m_{\rm h} \alpha}{2 \pi^{3/2} r_{\rm cut}}\frac{\exp[-(r/r_{\rm cut})^2]}{(r_{\rm c}^2+r^2)} \, ,
\end{equation}
where $m_{\rm h}$ is the halo mass, $r_{\rm c}$ is the core radius and
$\alpha \simeq 1.156$~\citep{2010MNRAS.408L..26P}.  The DM halo mass
can be estimated using the initial luminosity and a given
mass-to-light ratio.  Using the results from
\cite{NiedersteOstholt:2010hg} the initial luminosity is estimated to
be $\rm \sim 10^8 \, L_{\odot}$. Assuming a typical mass-to-light
ratio for dwarf galaxies of 25~\citep{1998ARA&A..36..435M}, the DM
halo mass is found to be $m_{\rm h} \,=\, \rm 2.4\times 10^9 \,
M_{\odot}$. To account for the initial tidal disruption of the SgrDw
halo by the Milky Way, a truncation of the halo profile is imposed at
$r_{\rm cut} \,=\, 12\, r_{\rm c}$. The evolution of the SgrDw in the Milky Way
potential is obtained via a N-body model of SgrDw using the particle-mesh gravity code SUPERBOX~\citep{2000NewA....5..305F}.
The evolution code allows to recover the actual DM profile by using the constraint of the observed stellar distribution.
The values of the parameters of the present ISO profile are given in Table~\ref{tab:jbarvalue}.

\section{Exclusion limits on the dark matter annihilation cross section \label{sec:limits}}

\begin{figure}[]
\begin{center}
\includegraphics[scale=0.5]{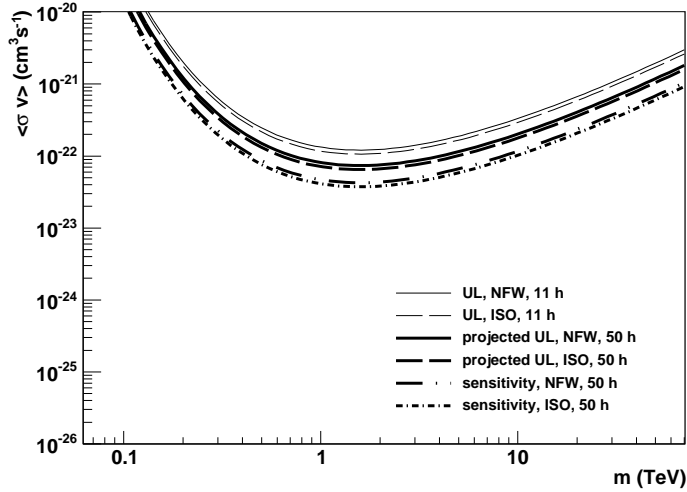}
\caption{95\% C.L. upper limits on the velocity-weighted annihilation
 cross section $\langle\sigma v\rangle$ versus the DM mass $m$ for a
 NFW (solid line) and Isothermal (ISO) (dashed line) DM halo profiles
 respectively for 11 h observation time and  $\rm \Delta\Omega\,=\,2\times10^{-5}\,sr$.
 The projected upper limits are displayed for 50 h observation time.
 The sensitivities at 95\% C.L. for 50 h are also shown for NFW (long-dashed dotted line) and ISO (dashed dotted line) DM halo profiles.}
\label{fig:exclusionlimits_HESS}
\end{center}
\end{figure}

\begin{figure}[]
\begin{center}
\includegraphics[scale=0.5]{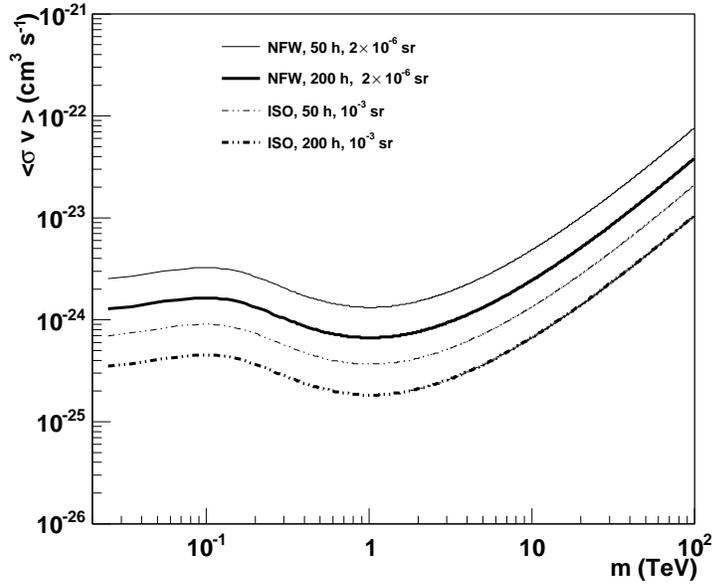}
\caption{Sensitivity at 95\% C.L. for CTA on the
  velocity-weighted annihilation cross section $\langle\sigma
  v\rangle$ versus the DM mass $m$ for a NFW (solid line) and
  Isothermal (ISO) (dashed line) DM halo profiles, respectively. The
  sensitivity is shown for 50 and 200 h observation times. The solid
  angle of observation is taken as $\rm
  \Delta\Omega\,=\,2\times10^{-6}\,sr$ for the NFW DM halo profile and
  $\rm \Delta\Omega\,=\,10^{-3}\,sr$ for the ISO DM halo profile.}
\label{fig:limits_CTA}
\end{center}
\end{figure}

Theories beyond the Standard Model (SM) of particle physics propose
several particle DM candidates. For instance, some supersymmetric
extensions of the SM predict a \emph{neutralino} as the lightest
stable supersymmetric particle, which is a good candidate for
DM~\citep{1996PhR...267..195J,2000RPPh...63..793B}. The parametrization of the neutralino
self-annihilation gamma-ray spectrum $\rmd N_{\gamma}$/$\rmd E_{\gamma}$
is taken from~\cite{1998APh.....9..137B} for a typical neutralino
annihilating into W and Z pairs. Fig.~\ref{fig:exclusionlimits_HESS}
shows the upper limits of current IACTs on $\langle \sigma v
\rangle$ as a function of the DM mass $m$ for $\rm
\Delta\Omega\,=\,2\,\times\,10^{-5}\,sr$. Using the HESS upper
limits published in \cite{2008APh....29...55A}, the new upper
limits are calculated for the NFW and ISO DM halo profiles of
Section~\ref{sec:modelling} and $11\Unit{h}$ of observation time; the projected upper limits
for $50\Unit{h}$ of observation time is also
plotted. The limits are at the level of $5\times10^{-23}\Unit{cm^3s^{-1}}$
around 1 TeV for $50\Unit{h}$. The sensitivity of H.E.S.S. for 50 h observation time is also displayed.
The sensitivity limits for CTA on
$\langle\sigma v\rangle$ as a function of the DM mass $m$ are
presented in Fig. \ref{fig:limits_CTA} for 50 h and 200 h observation
times. The limits are calculated with $\rm
\Delta\Omega\,=\,2\,\times\,10^{-6}\,sr$ for the NFW DM halo profile
and $\rm \Delta\Omega\,=\,10^{-3}\,sr$ for the ISO DM halo
profile. The sensitivity limits at 95\% C.L. reaches the level of $10^{-25}\Unit{cm^3s^{-1}}$ for DM masses of about 1 TeV in the case of the ISO DM halo profile.

\begin{figure}[]
\begin{center}
\includegraphics[scale=0.5]{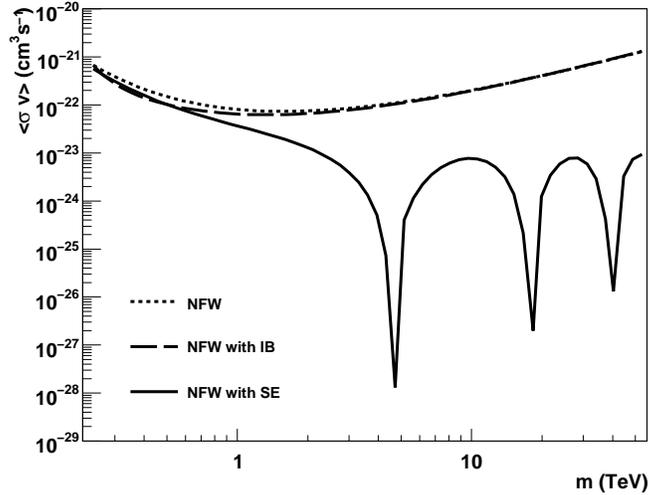}
\caption{Projected upper limits at 95\% C.L. on the $\langle\sigma v\rangle$/S
  versus the DM mass $m$ enhanced by the IB (dashed line) and SE (solid line) for the NFW
  profile. The projected upper limits are shown for 50 h
  observation times and $\rm \Delta\Omega\,=\,2\times10^{-5}\,sr$.}
\label{fig:exclusionlimits_HESS_IB_SE}
\end{center}
\end{figure}

\begin{figure}[]
\begin{center}
\includegraphics[scale=0.5]{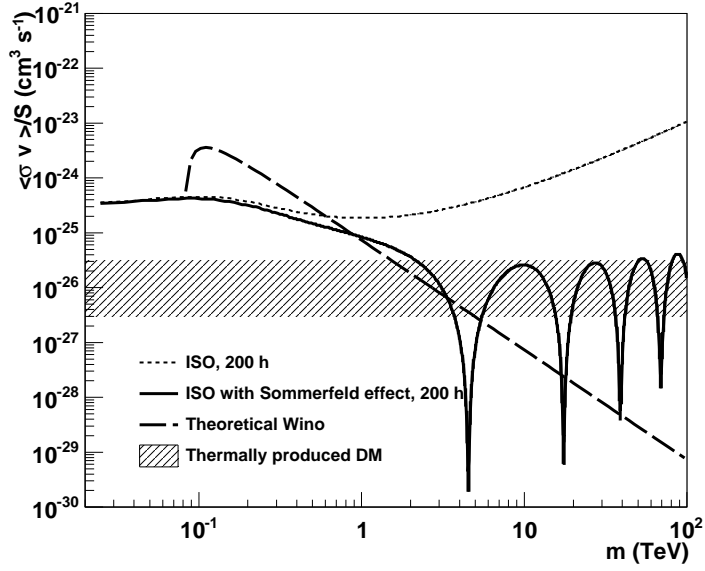}
\caption{Sensitivity at 95\% C.L. for CTA on the $\langle\sigma v\rangle$/S
versus the DM mass $m$ enhanced by the SE for the ISO profile. The sensitivity is
shown for 200 h observation times and $\rm \Delta\Omega\,=\,10^{-3}\,sr$.}
\label{fig:limits_IB_SOM_CTA}
\end{center}
\end{figure}

Two additional contributions to the overall gamma-ray flux that can
modify the limits are considered: namely the
\emph{Sommerfeld effect} and \emph{Internal Brems\-strahlung} (IB)
from the DM annihilation.  The Sommerfeld effect is a non-relativistic
effect which arises when two DM particles interact in an attractive
potential. When the relative velocity between the DM particles is
sufficiently low, the Sommerfeld effect can substantially boost the
annihilation cross section~\citep{Lattanzi:2008qa}, since it is
particularly effective in the very low-velocity regime. The actual
velocity-weighted annihilation cross section of the neutralino can
then be enhanced by a factor S defined as
\begin{equation}
\left\langle \sigma v \right\rangle = S\left\langle \sigma v \right\rangle_0 \, ,
\end{equation}
where the value of $S$ depends on the mass and relative velocity of
the DM particle. Assuming that the DM particles only annihilate to a W
boson, the attractive potential created by the Z gauge boson through
the weak force before annihilation would give rise to an
enhancement. Assuming that the DM velocity dispersion inside the halo
is the same as for the stars, the value of the DM velocity dispersion
is fixed at 11 kms$^{-1}$ for SgrDw~\citep{1998ARA&A..36..435M}. The value
of the enhancement is numerically calculated as done
in~\cite{Lattanzi:2008qa} and then used to improve the
upper limits on the velocity-weighted annihilation cross
section, $\left\langle \sigma v \right\rangle / S$ as a function of
the DM particle mass. Additionally, every time a DM particle
annihilates into charged particles, the electromagnetic radiative
correction to the main annihilation channel can give a more or less
significant enhancement to the expected gamma-ray flux in the observed
environment due to internal Bremsstrahlung
(IB)~\citep{Bergstrom:1989jr,Bringmann:2007nk}. Restraining the MSSM
parameters space to the \emph{stau co-annihilation region} of the
minimal supergravity (mSUGRA) models, for instance, the wino
annihilation spectrum would receive a considerable contribution from
Internal
Bremsstrahlung~\citep{Bringmann:2007nk}. Fig.~\ref{fig:exclusionlimits_HESS_IB_SE}
shows the 95\% C.L. upper limits on $\langle\sigma v\rangle / S$ as a
function of the DM mass $m$ for current IACTs. The projected upper limit
is shown for the NFW profile, $50\Unit{h}$ observation time and $\rm
\Delta\Omega\,=\,2\,\times\,10^{-5}\,sr$.  The effect of the IB is
only significant below $\sim 1 \Unit{TeV}$. Some specific wino masses can be
excluded due to the resonant enhancement in the Sommerfeld
effect. Outside resonances, the projected upper limits are improved by more
than one order of magnitude for DM masses above 1 TeV. The sensitivity at 95\% C.L. 
for CTA on $\langle\sigma v\rangle/S$ as a function of the DM
mass $m$ is presented in Fig.~\ref{fig:limits_IB_SOM_CTA}. The limits
are calculated for the ISO DM halo profile, with $200\Unit{h}$ observation
time and $\rm \Delta\Omega\,=\,10^{-3}\,sr$. The values of $\langle\sigma
v\rangle$ corresponding to cosmological thermally-produced DM, $\langle\sigma
v\rangle$ $\sim 3\times 10^{-26}$ cm$^3$s$^{-1}$, can be tested for specific wino masses in the resonance regions of the
Sommerfeld effect. Outside the resonances the sensitivity on $\langle\sigma
v\rangle / S $ is improved by more than one order of magnitude for
TeV DM masses, reaching the level of 10$^{-26}$ cm$^3$s$^{-1}$.

\section{Astrophysical background emission \label{sec:backgrounds}}

Dwarf galaxies are generally believed to contain very little
background emission from conventional astrophysical sources at VHE
energies, and are therefore easy targets for DM searches.  This
assumption is based on their low gas content and stellar formation
rate.  However, some gamma-ray emitting sources may still exist within
them: in particular from pulsars, and black hole accretion and/or jet emission processes.
The Sagittarius and Carina dwarf galaxies both host globular clusters
(the M54 globular cluster is located at the center of SgrDw), and
globular clusters are known to host millisecond pulsars (MSPs). The
collective emission of high energy gamma-rays by MSPs in globular
clusters has been detected by \Fermi~\citep{2010A+A...524A..75A},
and emission in the VHE energy range has been predicted by several
models for these objects, but has not yet been observed.  The possible
emission of very high energy radiation by millisecond pulsars from the
M54 globular cluster is examined in section \ref{sec:msp}.
Additionally, it has been suggested by some authors \citep[see][and
references thereby]{2007ApJ...668L.139L,2008ApJ...676.1008N} that
globular clusters may host black holes with masses of around $10^{2}$
to $10^{4}$ solar masses (called \emph{intermediate-mass black holes,
  or IMBHs}).  Indeed, \citet{2009ApJ...699L.169I} suggest SgrDw may
also be a possible host for a 10$^{4}$ M$_{\bigodot}$ IMBH . Their
claim is based on the study of the density profile around the central
point and the observed rise in the velocity dispersion of stars.  The
high energy emission from the IMBH candidate in the center of M54 is
discussed in section \ref{sec:imbh}

\subsection{Millisecond pulsars in M54}\label{sec:msp}

The M54 globular cluster at the center of SgrDw is likely to harbor a
large population of pulsars, especially MSPs.  The number of MSPs in
globular clusters has been shown by the \Fermi\ collaboration
\citep{2010A+A...524A..75A} to be correlated with the collision rate
$\Gamma$ defined by
\begin{equation}
\Gamma = \rho^{3/2}r_{\rm c}^{2} \, .
\end{equation}
In this equation, $\rho$ is the central luminosity and $r_{\rm c}$ is
the core radius. Taking a central surface brightness of $\rm \mu_{V}
\simeq (14.12\ - 14.9)\, mag \,arcsec^{-2}$ from Table 4 of
\cite{2008AJ....136.1147B} and a core radius $r_{c}=0.9\ \mbox{pc }$,
the collision rate is found to be
\begin{equation}
\Gamma_{\rm M54} \simeq \big(0.8-2.6\big)\times \Gamma_{\rm M62}\, ,
\label{eq:Gamma}
\end{equation}
where $\rm \Gamma_{\rm M62} = 6.5 \times 10^6\, L_{\odot}^{3/2}
{\mbox{pc}}^{-2.5}$ is the reference collision rate of the M62
globular cluster.
The predicted number $N_{\rm MSP}$ of MSPs in M54 is estimated from the collision rate \citep{2010A+A...524A..75A} by the relation
\begin{equation}
N_{\rm MSP} = 18+50\times\left(\frac{\Gamma_{\rm M54}}{\Gamma_{\rm M62}}\right)\, .
\label{eq:collisionrate}
\end{equation}
The collision rate from Eq.~(\ref{eq:Gamma}) gives the estimated
number of MSPs in M54: $N_{\rm MSP} = 60-140$. Note however that no
MSP has been discovered to date in M54.

The collective very-high-energy gamma-ray emission of millisecond
pulsars from globular clusters has been predicted by several authors,
notably Bednarek and Sitarek (BS) \citep{2007MNRAS.377..920B}, Venter,
deJager and Clapson (VJC) \citep{2009ApJ...696L..52V} and Cheng et
al. (CCDHK) \citep{2010ApJ...723.1219C}.
Using the effective area of CTA described in section
\ref{sec:currentinst}, one expects to observe respectively 1285 and
181 gamma-rays per hour towards the 47~Tucanae globular cluster, with
the BS and CCDHK models. In the latter model, the relic gamma-rays are
assumed to be the target population. The prediction of the VJC model
is somewhat smaller, only 71 gamma-rays per hour are predicted
assuming an interstellar magnetic field of 10 $\mu$G. The VJC model
also predicts a synchrotron radiation emission. The emission in the
keV range is predicted to be at the level of
$10^{-16}\Unit{TeV\,cm^{-2}\,s^{-1}}$ for a magnetic field of
$10\Unit{\mu G}$. This is easily accommodated by the measured diffuse
X-ray emission in M54 which is $\sim 2\times
10^{-14}\Unit{TeV\,cm^{-2}\,s^{-1}}$ \citep{2010A+A...512A..16B}.

As suggested by \citeauthor{2008AIPC.1085..277V} (\citeyear{2008AIPC.1085..277V}), a rough estimate of
the collective VHE emission of M54 can be obtained from their
predicted emission of 47~Tucanae by scaling by the factor:
\begin{equation}
x = \left(\frac{N_{\rm MSP}}{100}\right){\left(\frac{d_{\rm 47Tuc}}{d_{\rm M54}}\right)}^2{\left(\frac{<u_{\rm M54}>}{<u_{\rm 47Tuc}>}\right)}\, .
\end{equation}
In this equation, $d_{\rm 47Tuc}$ and $d_{\rm M54}$ are the distances
to 47~Tucanae and M54, and $<u_{\rm M54}>$ and $<u_{\rm 47Tuc}>$ the
average luminosity per cubic parsec of the globular cluster.
Taking the distances, luminosity and half-mass radii of M54 and
47~Tucanae from \cite{1996AJ....112.1487H} (2010 edition), one finds a
correction factor $x \simeq 1.6\times 10^{-2}$, assuming that M54
contains 100 MSPs. The expected number of gamma-rays per hour are thus
19.9 and 5.6 in the BS and CCDHK models. For the latter model, $x$ was
multiplied by an additional factor of 2 to take into account the
different number of MSPs in 47 Tucanae and M54. For the VJC model, the
number of expected gamma-rays per hour is about 1.1.

Whether this signal is observable or not depends crucially on its
spatial extension. The half-mass radius of M54 has an angular size of
less than $1'$ so that the signal would appear almost
point-like in the BS and VJC models. The CCDHK predicts an extended
signal. The electrons and positrons responsible for the
inverse Compton scattering on the CMB radiation have a typical
diffusion length of $100\Unit{pc}$, which corresponds to $\simeq
12\Unit{'}$ at the distance of M54. The signal integration regions are
taken as $3'$ for the BS and VJC models and $12'$ for the CCDHK model.
With an hadron rejection factor of 10\% as in section
\ref{sec:sensitivity}, the number of background per hour is $\sim$10
inside a $3'$ radius centered on M54. The significance of the
collective MSP signal depends thus on the observation time $T_{\rm
  obs}$ (in hours) as respectively $4.5\ \sqrt{T_{\rm obs}},$ $0.31\
\sqrt{T_{\rm obs}}$ and $0.25 \ \sqrt{T_{\rm obs}}$ in the BS, CCDHK
and VJC models. The BS model would give a signal at the $4.5\ \sigma$
level after just a one hour observation. The other models would give a much smaller signal, with a typical significance of $4 \sigma$ after 200 hours of observation.

In summary, the millisecond pulsars of M54 could give a significant VHE gamma-ray signal in CTA with observation times of typically 200 hours. For a cosmological thermally produced DM particle, $\langle\sigma v\rangle\, = 3\times$10$^{-26}$ cm$^3$s$^{-1}$, the corresponding signal would have a significance of 0.1$\sigma$, after 200 hours of observation and without any boost factor. The collective MSP signal would be a few orders of magnitude stronger than the DM annihilation signal.

\subsection{Intermediate-Mass Black hole}\label{sec:imbh}
%It has been suggested by some authors (see \citep{2007ApJ...668L.139L},\cite{2008ApJ...676.1008N} and references thereby) that globular clusters may host black holes with masses of around $10^{2}$ to $10^{4}$ solar masses, i.e. ``Intermediate Mass'' Black Holes or IMBH.
%The system composed by the dwarf galaxy Sgr and the Globular Cluster M54 has been recently studied by \cite{2009ApJ...699L.169I}  as a possible host for a 10$^{4}$ M$_{bigodot}$ IMBH. Their claim is based on the study of the density profile around the central point and the observed rise in the velocity dispersion of stars.
Significant radio and X-ray emissions are expected if the hypothesis
of a central IMBH is valid. Unfortunately, only upper limits to the
radio emission could be established from \textit{VLA} and
\textit{MOST} observations. Nevertheless, these limits can be used to
constrain the candidate black hole mass. As regards X-ray emission,
Ramsay and Wu~\citep{2006A+A...447..199R,2006A+A...459..777R} have
analyzed the data taken with \textit{Chandra} satellite. They found 7
bright sources within the half-mass radius of M 54. Their source
number 2 lies within 1'' of the density center of M54. Taking into
account the \textit{Chandra} astrometric accuracy of 0''.6 and the
systematics of 0.''3 in the absolute position of the \textit{Chandra}
ACS camera, this source could be associated with the stellar cusp
identified in \cite{2009ApJ...699L.169I}. The source number 2 has an
irregular shape and a luminosity of $\rm L_X = 0.72 \times 10^{33}\Unit{erg\,s^{-1}}$~\citep{2006A+A...447..199R,2006A+A...459..777R}.

\begin{figure}[]
\begin{center}
\includegraphics[scale=0.5,angle=270]{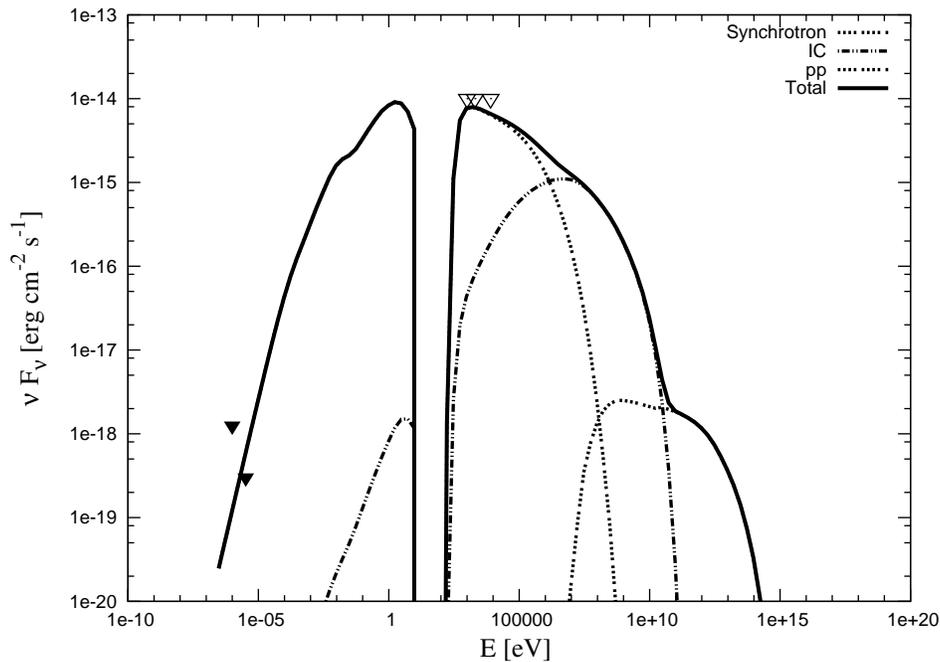}
\caption{Modeled emission of the candidate IMBH in M54.  The inverted
  empty triangles show the X ray emission from source 2 of Ramsay and
  Wu~(\citeyear{2006A+A...447..199R,2006A+A...459..777R}) and the inverted filled triangle show radio upper limits. The
  various contributions to the emission are shown. Only the $pp$
  emission contributes in the CTA energy range. The values of the
  parameters of the black hole model are displayed in Table~\ref{Tab:paramsgr}.}
\label{fig:IMBHflux}
\end{center}
\end{figure}

The \citet{2009ApJ...699L.169I} estimate of the black hole mass is
consistent with the (Massive Black Hole - host galaxy) correlation of
\citet{2006ApJ...644L..21F} only if the host system is M54 ($M_{\rm
  IMBH}/M_{\rm M54} \sim$ 5\%). For a similar mass ratio with SgrDw as
the host system, a 1000 times more massive black hole would be
necessary, suggesting this may be a system composed of a dwarf galaxy
hosting a prominent stellar nucleus, itself hosting a central IMBH.
We estimate the largest contribution of the IMBH to a possible VHE gamma-ray signal, and assume that the IMBH is active
and has a jet inclined towards the line of sight with an angle
$\theta.$ The contribution of the black hole to the VHE gamma-ray
emission is estimated using the model developed by
\cite{2011AA...531A..30R}, on the emission of relativistic jets
associated with active galactic nuclei. The parameters of the model
for the central black hole and jet are described
in~\cite{2011AA...531A..30R}. The calculation also uses the
constraints from the upper limits in the radio band and the measured
X-ray emission from the source number 2.
The modeled gamma-ray emission is shown on Fig.~\ref{fig:IMBHflux}. The parameters used in the model are given in
Table~\ref{Tab:paramsgr}. The emission depends only weakly on the
black hole mass, but strongly on the assumed Lorentz factor
$\Gamma_{\rm b}$ and inclination $\theta.$

The peak on the X-ray band comes from the synchrotron of electrons
while a strong contribution from the Synchrotron Self-Compton (SSC) scattering can be seen at GeV
energies. At higher energies, in particular in the CTA energy range,
the emission from $pp$ interactions is dominant. However, for
reasonable parameters, it is in the $10^{-18}-10^{-17}\Unit{erg
  \,cm^{-2}\,s^{-1}}$ flux range---too faint to be detected by CTA.

\section{Summary \label{sec:summary}}

Older publications \citep[e.g.][]{Evans:2003sc, 2008APh....29...55A}
on DM searches towards SgrDw used dark matter mass profiles which lead to
somewhat optimistic constraints on particle dark matter
self-annihilation cross sections. These models were used because no
accurate modelling of SgrDw existed at that time. Several realistic
models are now published that loosen the existing constraints by more
than one order of magnitude. The future CTA array will be sensitive to
$\langle\sigma v\rangle$ values around a few
$10^{-25}\Unit{cm^{3}\,s^{-1}}$. Some models could be excluded after
200 hours of observation, if boosts factors are taken into account.

However, the very high energy emission of several astrophysical
objects could give an observable signal for long-enough observation
times. The collective very high energy emission of the MSPs of the M54
globular cluster, which is predicted by several models, could be much
stronger than a DM signal.  It could be observed in just a few tens of
hours with CTA.  The candidate IMBH located at the center is not
expected to give an observable signal.  Under favorable circumstances
(active black hole and jet aligned towards the line of sight), it
might nevertheless be detectable in observations of SgrDw.

\newpage

{\small
\begin{table}[!ht]
\begin{center}
  \caption{Values of the $LOS$-integrated squared density averaged
    over the solid angle ($\bar{J}$) expressed in units of
    $10^{23}\Unit{GeV^2\,cm^{-5}}$, for different solid angles $\rm
    \Delta\Omega$. The values of $\Jbar$ are calculated for the NFW
    and ISO DM halo profiles. The parameters of these profiles are
    given in the first column. \label{tab:jbarvalue}}
\begin{tabular}{cccc}
\\
\hline\hline
DM halo profile & $\rm \Delta\Omega = 10^{-3}$ sr &$\rm \Delta\Omega = 2\times10^{-5}$ sr &  $\rm \Delta\Omega = 2\times 10^{-6}$ sr  \\
\hline\hline
NFW & 0.065 & 0.88& 3.0 \\
 $r_{\rm s}$ = 1.3 kpc  & & & \\
 $\rho_{\rm s}$  = 1.1 $\rm \times 10^{-2} \, M_{\odot} pc^{-3}$&&&\\
\hline
ISO &   0.49 & 1.0& 1.0\\
  $r_{\rm c}$ = 0.34 kpc &  &  &\\
$m_{\rm h}$  = $\rm 9.5\times 10^8\,M_{\odot}$ &&& \\

\hline\hline
\end{tabular}
\end{center}
\end{table}
}

   \begin{table}[!htp]
      \caption[]{Model parameters for the IMBH candidate in M54.}
         \label{Tab:paramsgr}
     $$
         \begin{array}{p{0.75\linewidth}l}
            \hline
            \noalign{\smallskip}
            Parameter      &  {\rm Value}  \\
            \noalign{\smallskip}
            \hline
            \noalign{\smallskip}
            $M_{\rm bh}$: black hole mass & 5\times10^{3} \rm M_{\odot}\\
            $R_{\rm g}$: gravitational radius & 7.38\times 10^{8} {\rm cm} \\
            $L_{\rm j}^{\rm(kin)}$: jet kinetic power at $z_0$ &  6.28\times 10^{39} {\rm erg \ s^{-1}}   \\
           $q_{\rm j}$: ratio $2 L_{\rm j}^{\rm (kin)}/L_{\rm Edd}$ & 0.05 \\
            $\Gamma_{\rm b}(z_0)$:  bulk Lorentz factor of the jet at $z_0$  & 4   \\
%           Yorke 1979, Yorke 1980a & \leq 1700
            $\theta$: viewing angle &  45^\circ     \\
            $\xi_{\rm j}$: jet's half-opening angle &  5^\circ \\
            $q_{\rm rel}$: jet's content of relativistic particles & 0.05 \\
            $a$: hadron-to-lepton power ratio     & 1          \\
            $z_0$: jet's launching point   & 50 \ R_{\rm g} \\
            $s$: spectral index injection & 2.1\\
            $\eta$: acceleration efficiency & 1. \times 10^{-2}\\

            $N_H$: column dust density & 10^{21} \,{\rm cm^{-2}}\\

            \noalign{\smallskip}
            \hline
         \end{array}
     $$
   \end{table}

%-------------------------------------------------------------
%                      Acknowledgments
%-------------------------------------------------------------

\clearpage

\bibliographystyle{natbib}
\bibliography{SagittariusCTA}

\end{document}